\documentclass{PoS}

\def\gsim{\mathrel{\rlap{\lower4pt\hbox{\hskip1pt$\sim$}}
    \raise1pt\hbox{$>$}}}         %greater than or approx. symbol
\def\lsim{\mathrel{\rlap{\lower4pt\hbox{\hskip1pt$\sim$}}
    \raise1pt\hbox{$<$}}}         %less than or approx. symbol

\newcommand{\be}{\begin{equation}}
\newcommand{\ee}{\end{equation}}
\newcommand{\bea}{\begin{eqnarray}}
\newcommand{\eea}{\end{eqnarray}}
\newcommand{\bi}{\begin{itemize}}
\newcommand{\ei}{\end{itemize}}
\newcommand{\ben}{\begin{enumerate}}
\newcommand{\een}{\end{enumerate}}
\newcommand{\la}{\left\langle}
\newcommand{\ra}{\right\rangle}
\newcommand{\lc}{\left[}
\newcommand{\rc}{\right]}

\title{Improving quark flavor separation with forward $W$ and $Z$ production at LHCb}

\ShortTitle{Quark flavor separation with forward $W$ and $Z$ production at LHCb}

\author{\speaker{Juan Rojo}\thanks{On behalf of the NNPDF Collaboration.}\\
  Department of Physics and Astronomy,
VU University, De Boelelaan 1081,
1081 HV Amsterdam,\\ 
and Nikhef Theory Group,
Science Park 105, 1098 XG Amsterdam,
The Netherlands. \\
        E-mail: \email{j.rojo@vu.nl}}

\abstract{
  We quantify the constraints on the 
  flavour separation between the quarks
  and antiquarks in the proton
  provided by the recent forward weak gauge boson production
  data from the LHCb experiment at $\sqrt{s}=7$ and 8 TeV.
  Performed in the framework of the NNPDF3.1 global
  analysis, this study
  highlights the key role that the LHCb $W$ and $Z$ 
  data have in achieving a robust quark flavor separation
  in the large-$x$ region, including
  the strange and charm quarks. 
  We demonstrate how the LHCb measurements lead to improved
  determinations of the the up and down
  quark PDFs in the region $x\gsim 0.1$,
  with an uncertainty reduction that can be as large as
   a factor 2.
   We also show how the LHCb forward measurements
   severely restrict the size of the
   fitted charm PDF at large $x$, imposing stringent constraints
   on non-perturbative
  models for the charm content of the nucleon.
}

\FullConference{XXV International Workshop on Deep-Inelastic Scattering and Related Subjects\\
		3-7 April 2017\\
		University of Birmingham, UK}

\begin{document}

\vspace{0.1cm}
\noindent
The recent NNPDF3.1 global analysis~\cite{nnpdf31} is the latest
release from the NNPDF collaboration.
It is based on the NNPDF3.0 framework~\cite{Ball:2014uwa} with two main
improvements.
On the one hand, a large number of new precision collider measurements have been
added, including some that for the first time are included in a global
fit such as the $Z$ transverse momentum~\cite{Boughezal:2017nla}
and the differential distributions
in top-quark pair production~\cite{Czakon:2016olj}.
On the other hand, the charm PDF is now treated on an equal footing as
compared of the light quark
PDFs~\cite{Ball:2016neh}, that is, in NNPDF3.1 $c(x,Q_0)$ is parametrized
with an artificial neural network
and then fitted to the data.
This improved treatment of the charm PDF
has several benefits, among which that of reducing
the dependence of high-scale observables
to the value of the charm quark mass $m_c$.

Concerning the new experimental data, a wealth of recent electroweak gauge boson production measurements
has been included in NNPDF3.1.
These datasets
provide useful information
on the separation between different quark flavours, including
strange and charm, as well as between quarks and antiquarks.
In this respect, the forward weak boson production measurements
presented by the LHCb collaboration are particularly interesting to access the
large-$x$ region~\cite{Thorne:2008am}, where PDF uncertainties are
the largest.
NNPDF3.1 includes the
complete LHCb $\sqrt{s}=7$ and 8 TeV measurements of inclusive $W$ and $Z$
production in the forward region in the muon
channel~\cite{Aaij:2015gna,Aaij:2015zlq}, which supersede all previous
measurements in the same final state.
These new measurements complement earlier LHCb data already included in
NNPDF3.0~\cite{Aaij:2012vn,Aaij:2012mda}.
In this contribution, we explore the impact of the LHCb data on the
NNPDF3.1 analysis and show how sizable constraints are derived
at large-$x$
for all the light quark PDFs and for the charm PDF.
In turn, this leads to reduced theoretical
uncertainties for the production of new massive BSM resonances,
which are dominated
currently by  the large-$x$ PDF uncertainties~\cite{Beenakker:2015rna}.

%%%%%%%%%%%%%%%%%%%%%%%%%%%%%%%%%%%%%%%%%%%%%%%%%%%%%%%%%%%%%
\begin{figure}[t]
\centering
  \includegraphics[width=.74\linewidth]{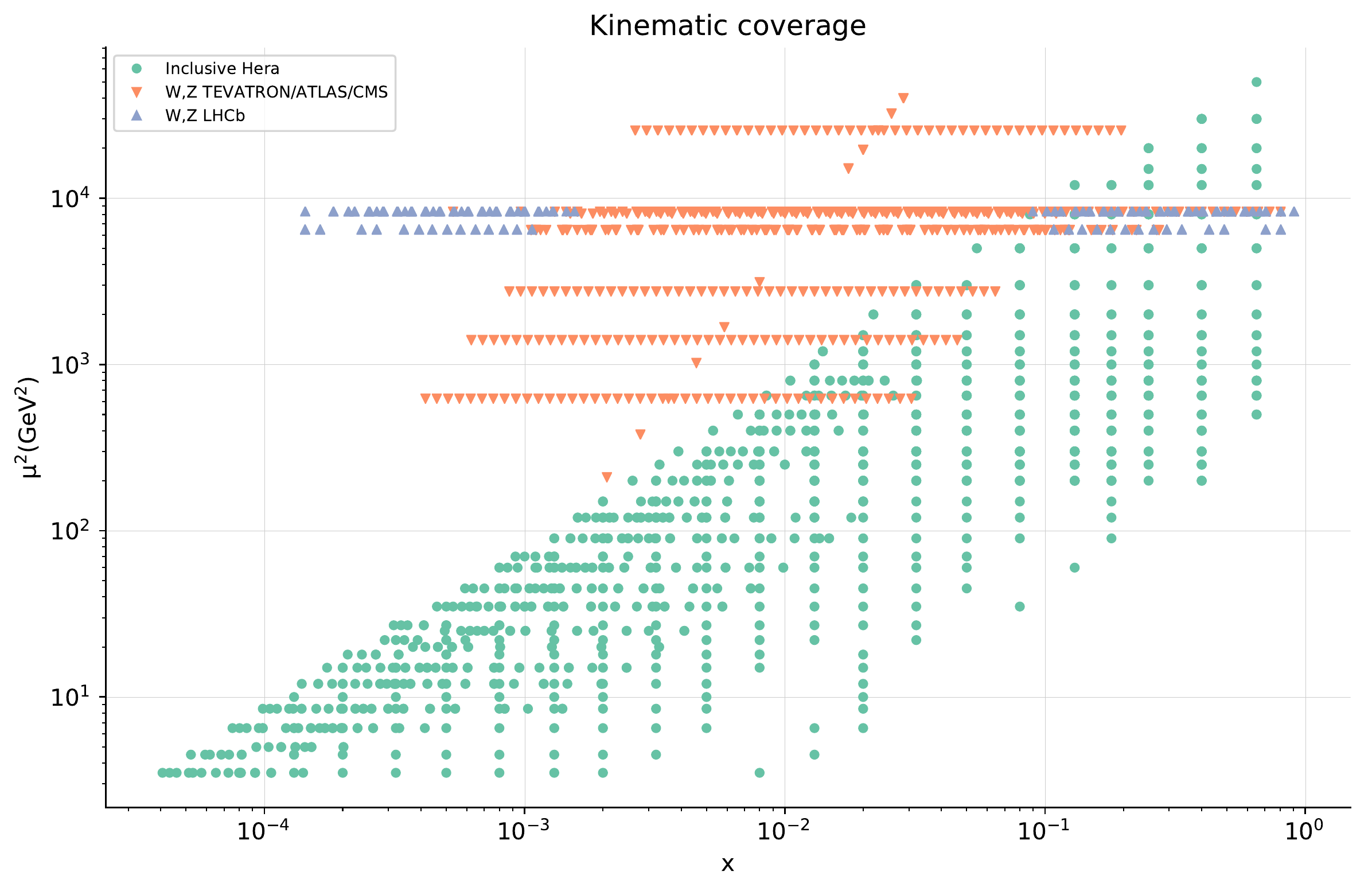}
  \caption{\small The coverage in the $(x,Q^2)$ plane of
    the $W,Z$ production data from LHCb, compared with that
     of $W,Z$ production at ATLAS, CMS and the Tevatron
   and with that of the HERA structure functions.
  }
\label{fig:kinplot}
\end{figure}
%%%%%%%%%%%%%%%%%%%%%%%%%%%%%%%%%%%%%%%%%%%%%%%%%%%%%%%%%%%%%%%%%%

Let us also mention that another set of recent LHCb measurements that have revealed an
unexpected usefulness for PDF studies is forward $D$ meson production~\cite{Zenaiev:2015rfa,Gauld:2015yia}.
As demonstrated in~\cite{Gauld:2016kpd}, the combination of LHCb charm production data at
$\sqrt{s}=5,7$ and 13 TeV leads to a reduction of the PDF uncertainties
on the gluon at $x\simeq 10^{-6}$ by up to an order of magnitude.

\vspace{0.1cm}
\noindent
{\bf Forward weak boson production at LHCb.}
In Fig.~\ref{fig:kinplot} we show the kinematical coverage in the $(x,Q^2)$
plane of the
$W$ and $Z$ production data from LHCb included
in NNPDF3.1, compared with the
    corresponding coverage of $W,Z$ production at ATLAS, CMS and the Tevatron,
    as well as with that of the HERA inclusive structure function data.
    For the Drell-Yan data, the values of $(x,Q^2)$ for each data bin
    are approximated assuming leading order kinematics, namely $x_{1,2}=(M/\sqrt{s})e^{\pm y}$,
    with $M$ and $y$ the invariant mass and the lepton rapidity of each bin.
    We observe that the LHCb data spans a wider and complementary range
    in $x$ as compared from the ATLAS and CMS measurements, and in particular
    ensure an improved coverage of the large-$x$ region.

\begin{figure}[t]
\centering
\includegraphics[width=.49\linewidth]{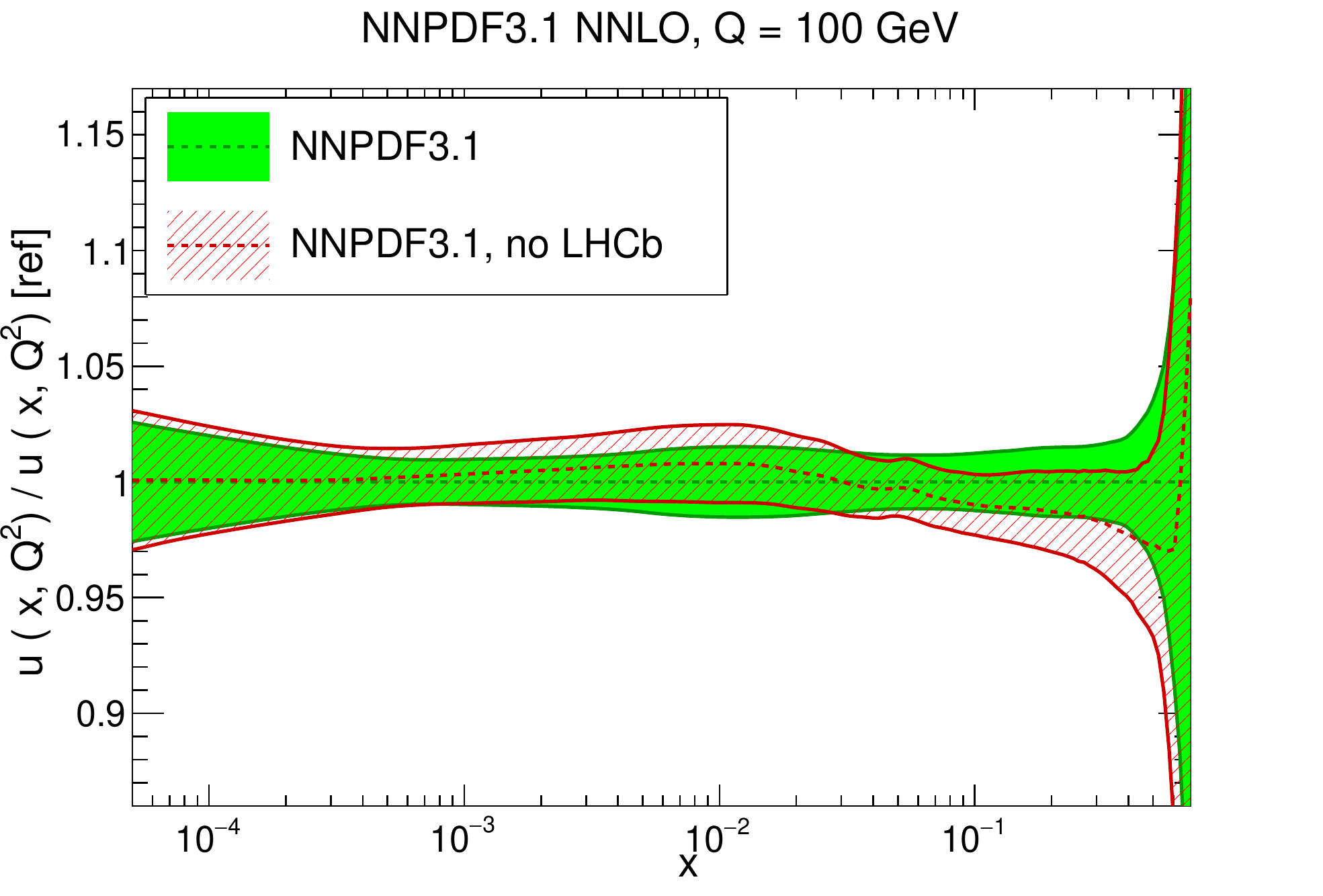}
\includegraphics[width=.49\linewidth]{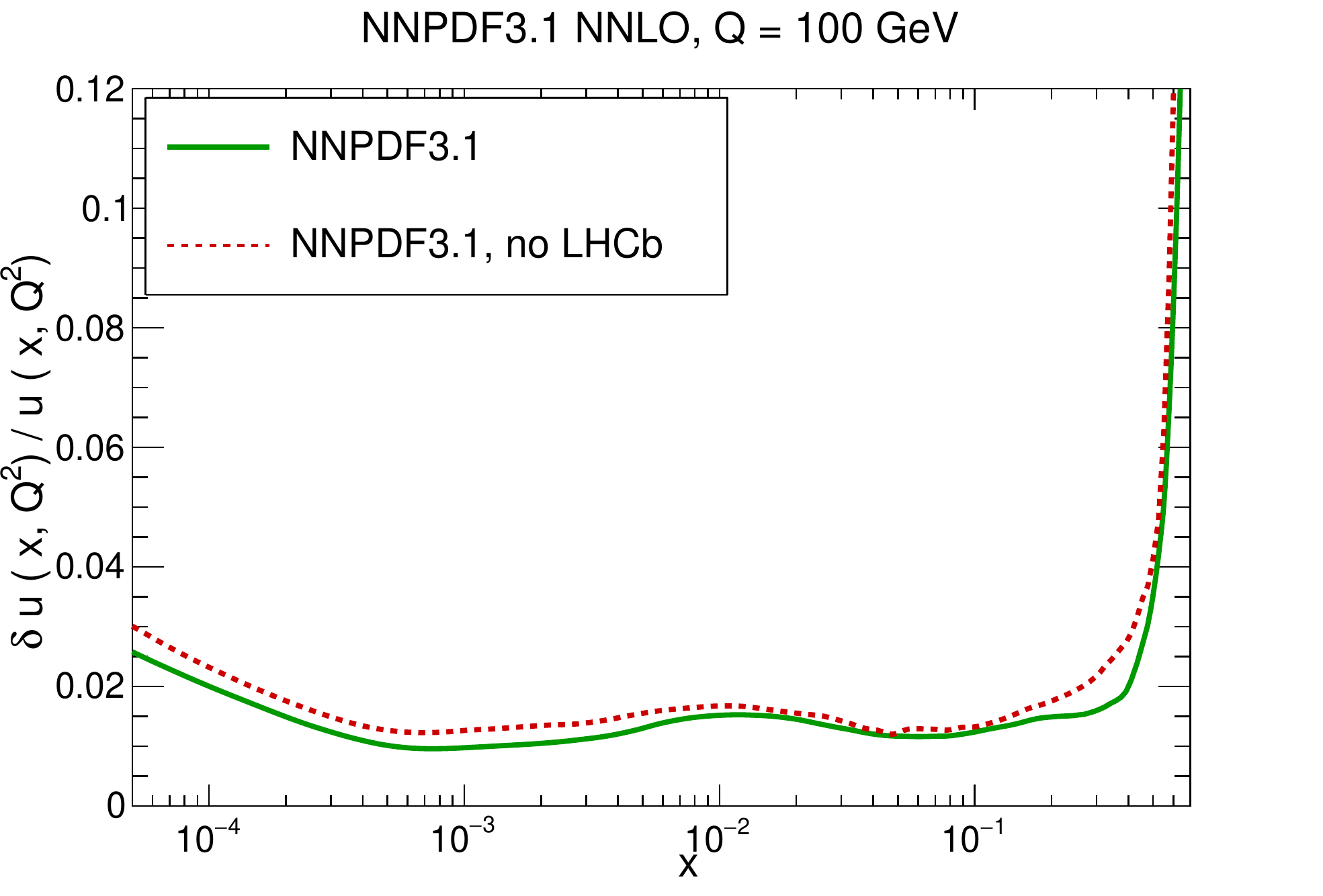}
\includegraphics[width=.49\linewidth]{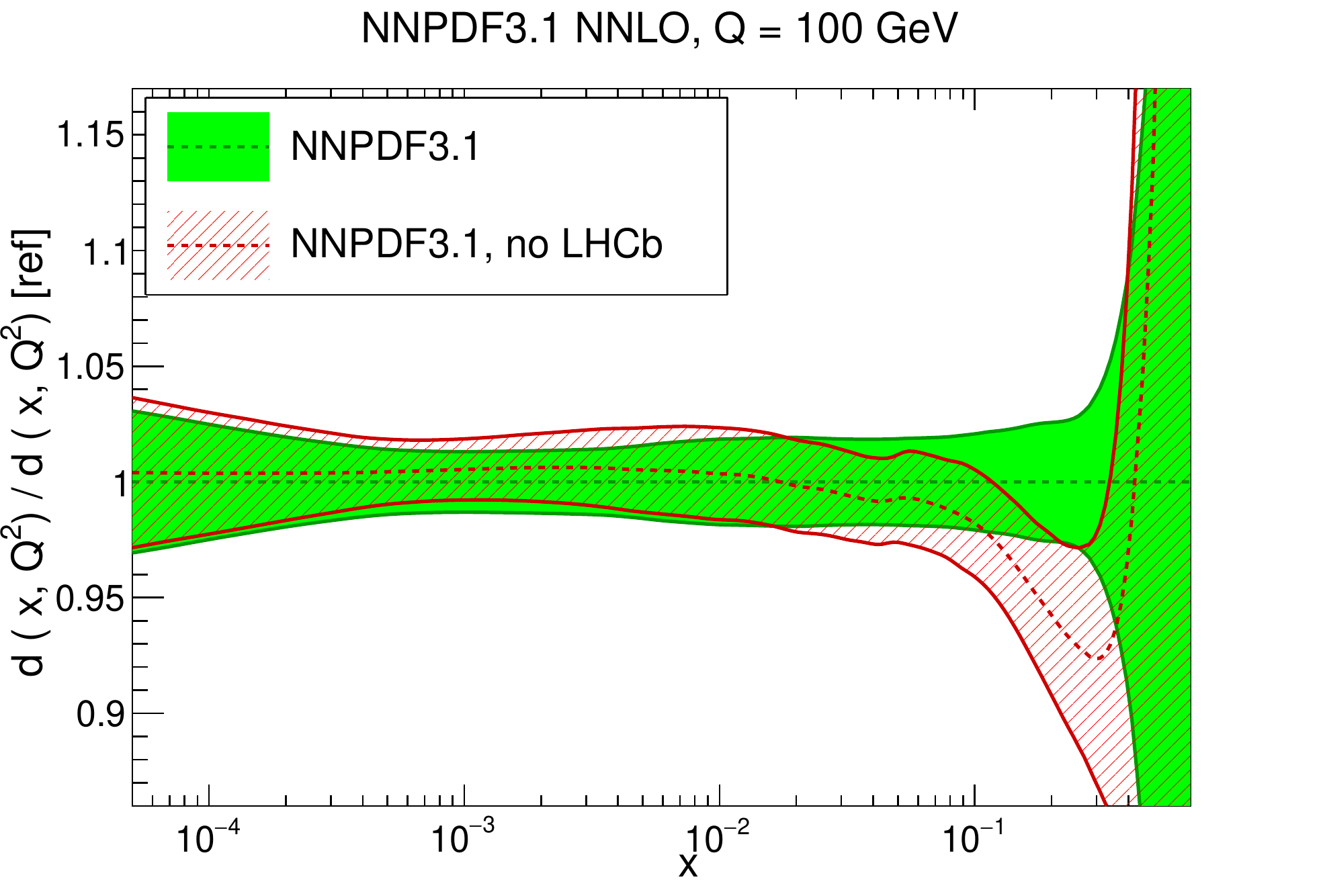}
 \includegraphics[width=.49\linewidth]{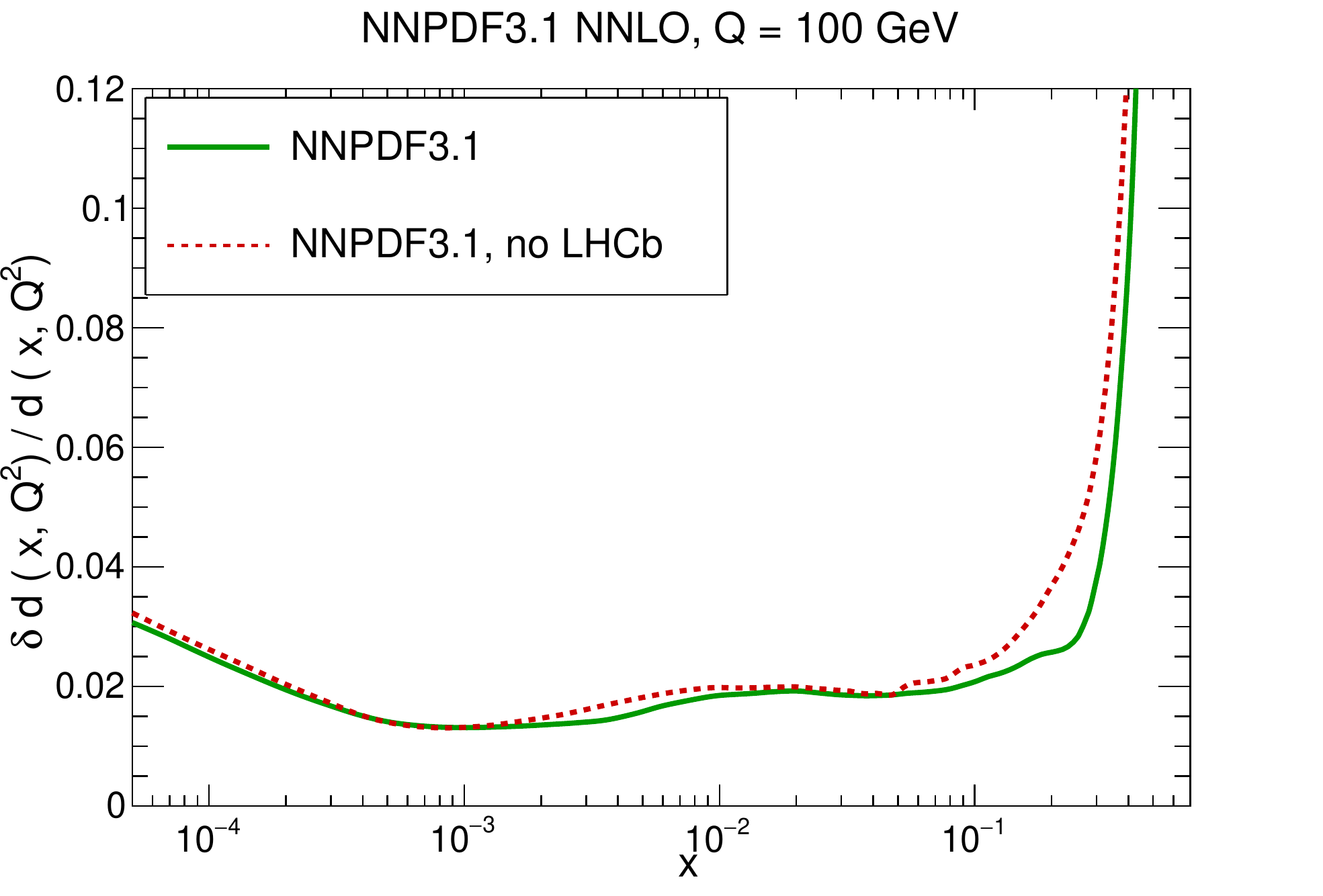}
 \caption{\small The up quark (upper plots) and down quark (lower plots)
   PDFs at $Q=100$ GeV, comparing the results of the NNPDF3.1 baseline with those
   of the corresponding fit without LHCb data.
   We show the PDF ratios normalized to the central value of NNPDF3.1 (left)
   and the relative PDF uncertainties (right plots).
  }
\label{fig:PDFcomp}
\end{figure}

    In order to illustrate the impact of the LHCb data,
in Fig.~\ref{fig:PDFcomp} we show the up quark  and down quark 
   PDFs at $Q=100$ GeV, comparing the results of the NNPDF3.1 NNLO fit  with those
   of the same fit without any LHCb data.
   We show the PDF ratios normalized to the central value of NNPDF3.1 
   and the relative PDF uncertainties.
   From this comparison, we see that the LHCb data has a significant
   impact in NNPDF3.1, both in terms of shifting the
   central value of the large-$x$ quarks, where the LHCb data prefer
   larger
   values, and in terms of reducing the PDF uncertainties.
   In the case of $xd(x,Q)$, the LHCb data reduce
   the PDF uncertainties by almost a factor 2 for $x\simeq 0.3$.
     
%%%%%%%%%%%%%%%%%%%%%%%%%%%%%%%%%%%%%%%%%%%%%%%%%%%%%%%%%%%%%
%%%%%%%%%%%%%%%%%%%%%%%%%%%%%%%%%%%%%%%%%%%%%%%%%%%%%%%%%%%%%%%%%%

\begin{figure}[t]
  \centering
  \includegraphics[width=.49\linewidth]{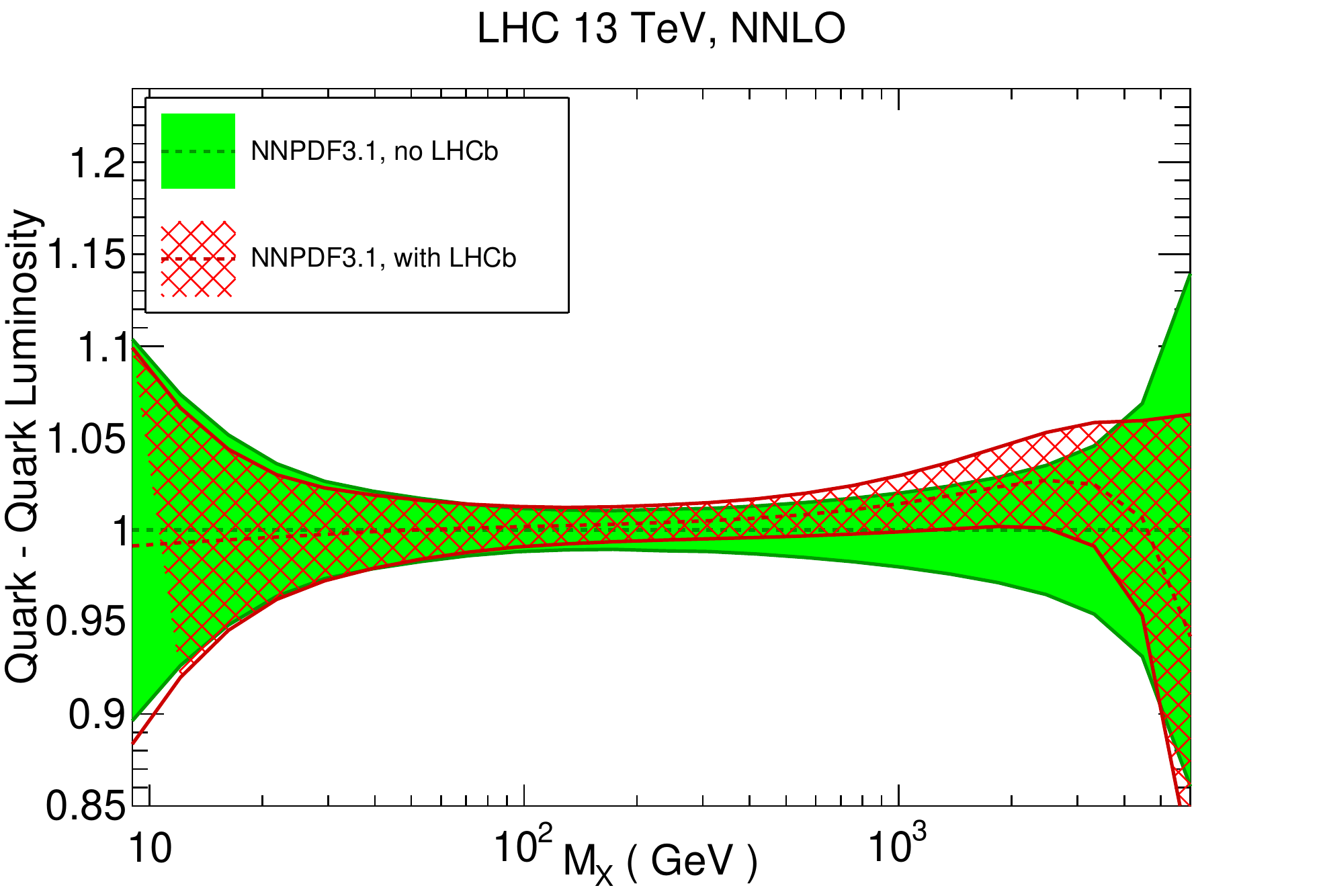}
\includegraphics[width=.49\linewidth]{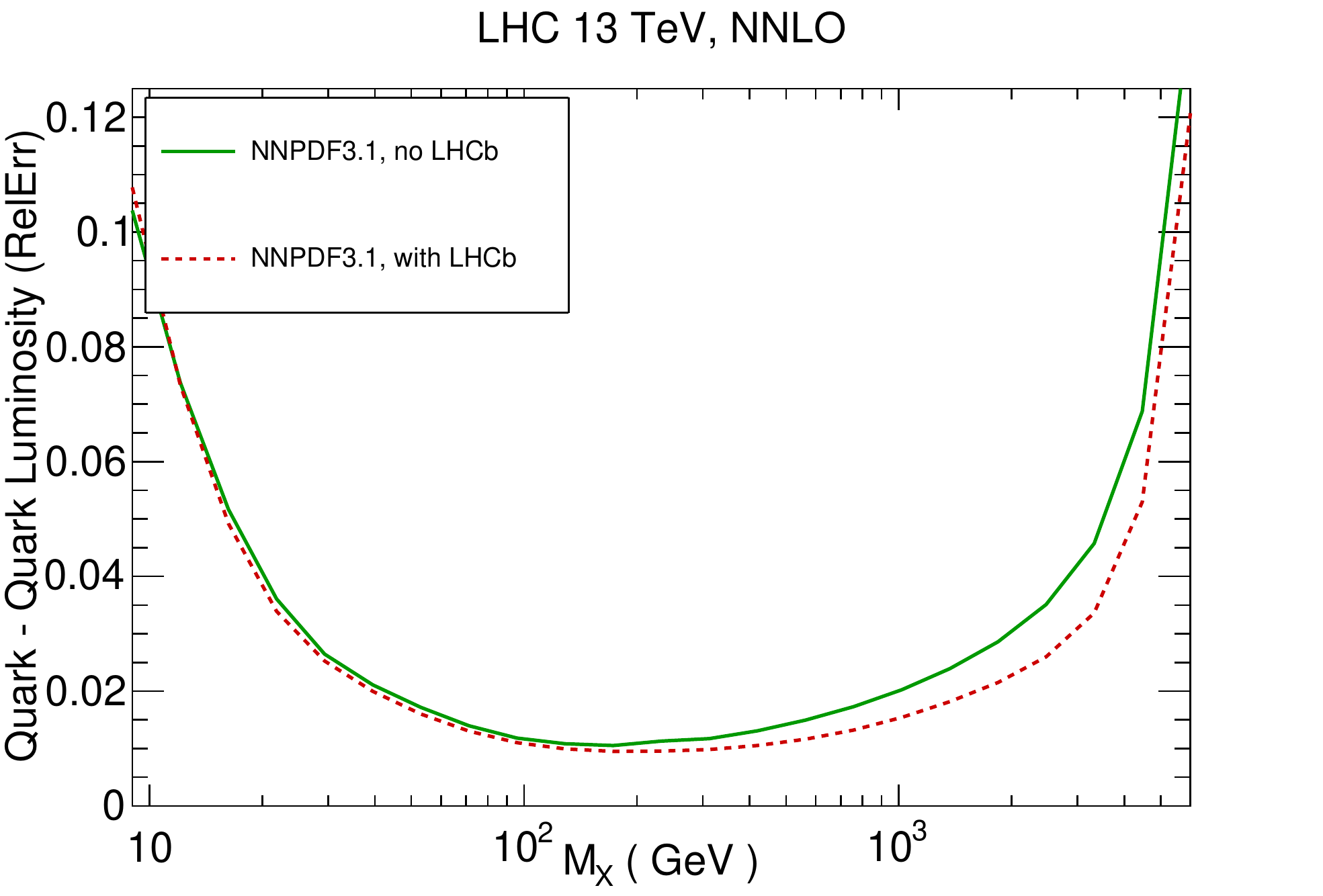}
\caption{\small The quark-quark PDF luminosity (left plot)
  and its relative uncertainty (right plot) comparing the
  NNPDF3.1 fits with and without the LHCb data included.
  }
\label{fig:PDFlumis}
\end{figure}

Next, in Fig.~\ref{fig:PDFlumis} we show 
 the quark-quark PDF luminosity $\mathcal{L}_{qq}$
  and its relative uncertainty for the
  NNPDF3.1 fits with and without the LHCb data.
  We find that the LHCb data prefers harder $\mathcal{L}_{qq}$
  values in the TeV region, and also that it leads to an important
  reduction of the PDF uncertainties in the same region.
  
%%%%%%%%%%%%%%%%%%%%%%%%%%%%%%%%%%%%%%%%%%%%%%%%%%%%%%%%%%%%%
%%%%%%%%%%%%%%%%%%%%%%%%%%%%%%%%%%%%%%%%%%%%%%%%%%%%%%%%%%%%%%%%%%

In order to quantify the improvement in the description of
the LHCb data achieved in the NNPDF3.1 analysis, in
Table~\ref{table:chi2} we show the
$\chi^2/N_{\rm dat}$ values for the individual LHCb datasets included in
    NNPDF3.1 (as well as their total) for three NNLO different fits:
    NNPDF3.1, NNPDF3.1 no LHCb, and NNPDF3.0.
    The numbers in brackets indicate those datasets not included in
    the corresponding fits.
   We see that NNPDF3.1 achieves a good description of the four LHCb datasets, markedly
   improving on
   the NNPDF3.0 predictions for the new 7 and 8 TeV measurements.
   We have also verified that the  use of
   NNLO theory is crucial to achieve satisfactory $\chi^2/N_{\rm dat}$ values.
   
%%%%%%%%%%%%%%%%%%%%%%%%%%%%%%%%%%%%%%%%%
\begin{table}[h]
  \centering
  \small 
  \begin{tabular}{|l|c|c|c|}
    \hline
  Dataset  &   NNPDF3.1   & NNPDF3.1 no LHCb  & NNPDF3.0 \\  
  \hline
  \hline
  LHCb $Z$ 940 pb  & 1.4   &  $\lc 1.7\rc$ &       1.3 \\
  LHCb $Z\to ee$ 2 fb  & 1.1  & $\lc 1.0 \rc$   & 1.2 \\
  LHCb $W,Z \to \mu$ 7 TeV &  1.7  & $\lc 4.4 \rc$  & $ \lc 2.6\rc $ \\
  LHCb $W,Z \to \mu$ 8 TeV &  1.4   & $\lc 3.4 \rc$  & $\lc 2.4 \rc $ \\
  \hline
  LHCb total  &  {\bf  1.5}   &  $\bf \lc 3.1\rc$    &   {\bf 2.1}  \\
  \hline
  \end{tabular}
  \caption{\small \label{table:chi2}
    The $\chi^2/N_{\rm dat}$ for the individual LHCb datasets included in
    NNPDF3.1 (as well as their total) for three different NNLO fits.
    The numbers in brackets indicate datasets not included in
    the corresponding fit.
  }
\end{table}
%%%%%%%%%%%%%%%%%%%%%%%%%%%%%%%%%%%%%%%

\begin{figure}[t]
  \centering
  \includegraphics[width=.99\linewidth]{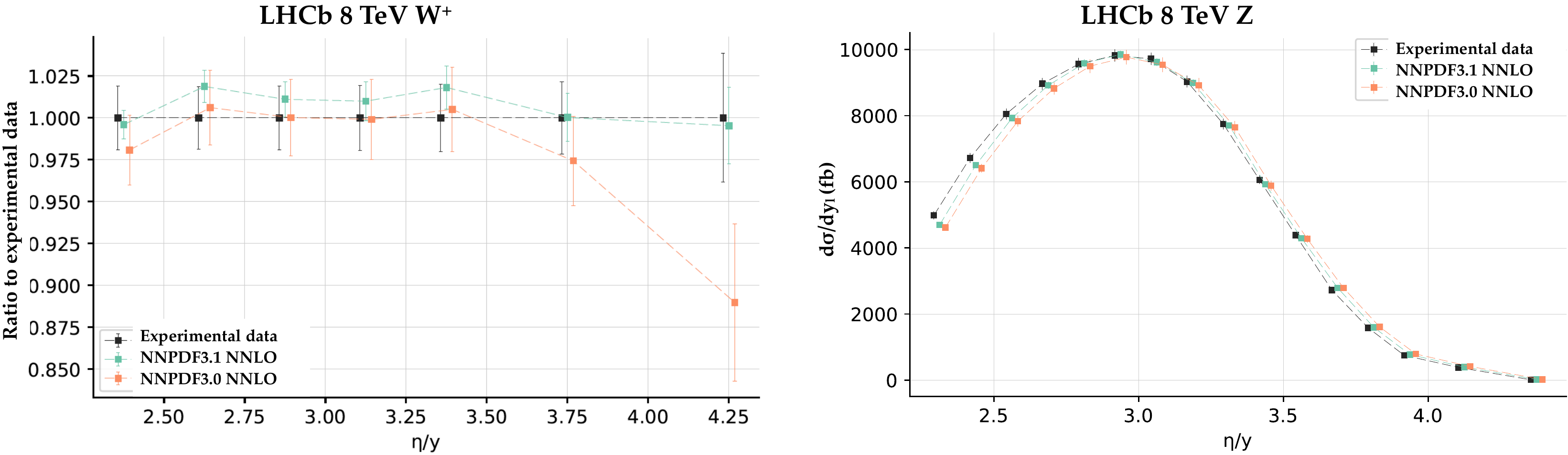}
  \caption{\small Left plot: comparison between the NNPDF3.1 and NNPDF3.0 NNLO predictions
    and the LHCb data on $W^+$ production at 8 TeV in the muon final state, normalized
    to the central value of the experimental data.
    Right plot: same comparison for the $Z$ rapidity distributuon, now
    in terms of the absolute cross-section.
  }
\label{fig:datath}
\end{figure}

To further illustrate the improved agreement between theory and data,
in Fig.~\ref{fig:datath} we show the comparison between the NNPDF3.1 and NNPDF3.0 NNLO predictions
and the LHCb data on the lepton rapidity distribution
in $W^+$ production at 8 TeV in the muon final state, normalized
    to the central value of the experimental data.
    We observe that not only the agreement is improved at the level of
    central values, specially in the forward region, but also that the
    PDF uncertainties for individual bins are reduced by up to a factor two
    in NNPDF3.1.
In the same figure we also show the corresponding comparison for the $Z$ rapidity distributuon, now
    in terms of the absolute cross-section.
    Also in this case we observe how the central values of the NNPDF3.1 prediction move
    closer to the data as compared to the NNPDF3.0 ones.
    It is also worth emphasizing that for all the LHCb
    datasets the agreement between
    NNLO theory and data is markedly better than at NLO.

%%%%%%%%%%%%%%%%%%%%%%%%%%%%%%%%%%%%%%%%%%%%%%%%%%%%%%%%%%%%%
%%%%%%%%%%%%%%%%%%%%%%%%%%%%%%%%%%%%%%%%%%%%%%%%%%%%%%%%%%%%%%%%%%

\vspace{0.1cm}
\noindent
{\bf Constraining the charm content of the proton.}
In the NNPDF3.1 global analysis, the  charm PDF is parametrized at the
input scale $Q_0 \gsim \mu_c$, with $\mu_c=m_c$ being
the charm threshold, and then determined from experimental data
in the same way as the light
quark PDFs.
It can be shown that
the forward $W,Z$ production data from LHCb provide, in addition
to constrains on the light quark PDFs, also useful
information on the charm content of the proton.
This sensitivity can be understood from the fact that
forward gauge boson production depends on charm PDF via 
partonic subprocesses such as $\bar{s}c\to W^+$ and $s\bar{c}\to W^-$.

\begin{figure}[t]
\centering
\includegraphics[width=.49\linewidth]{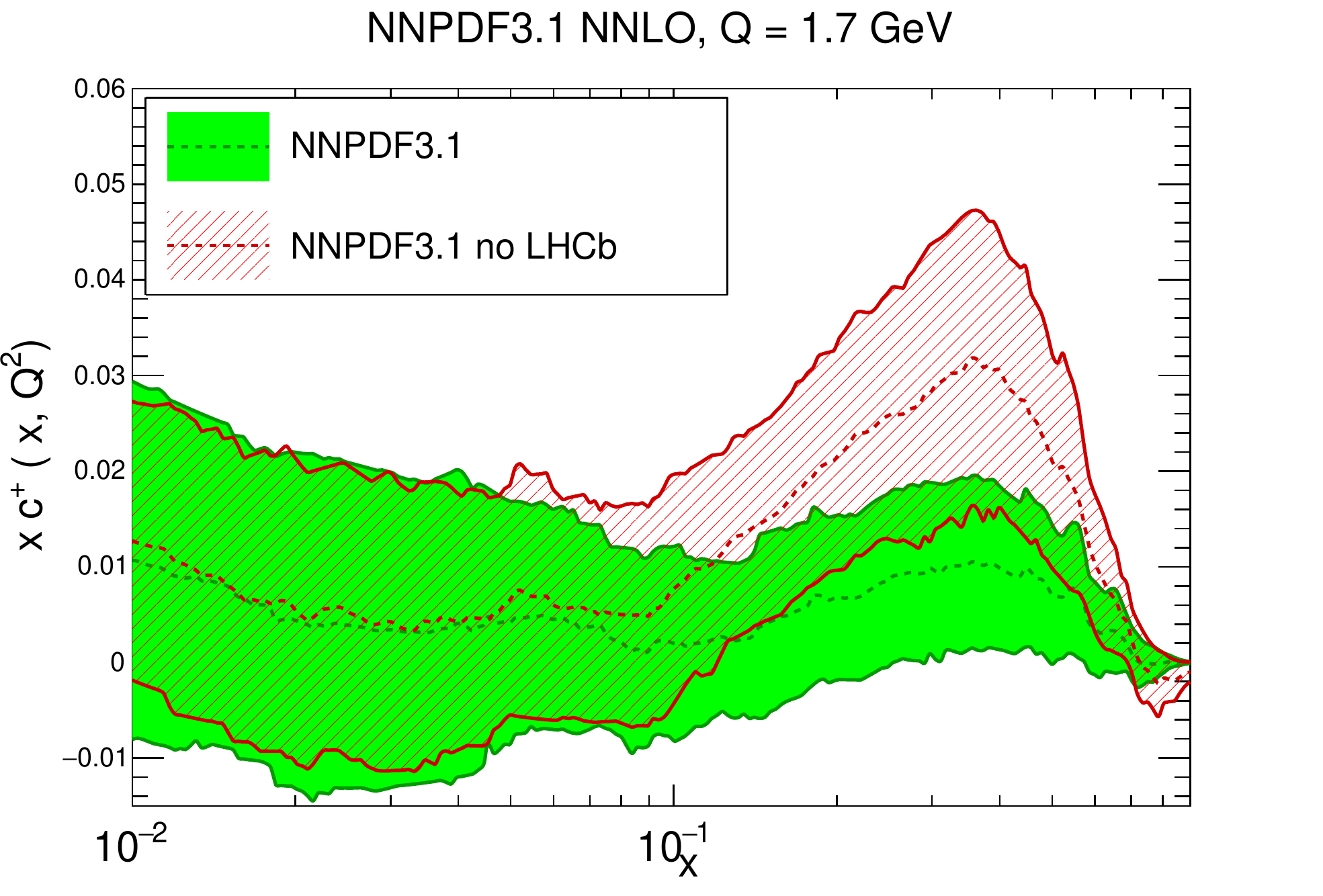}
\includegraphics[width=.49\linewidth]{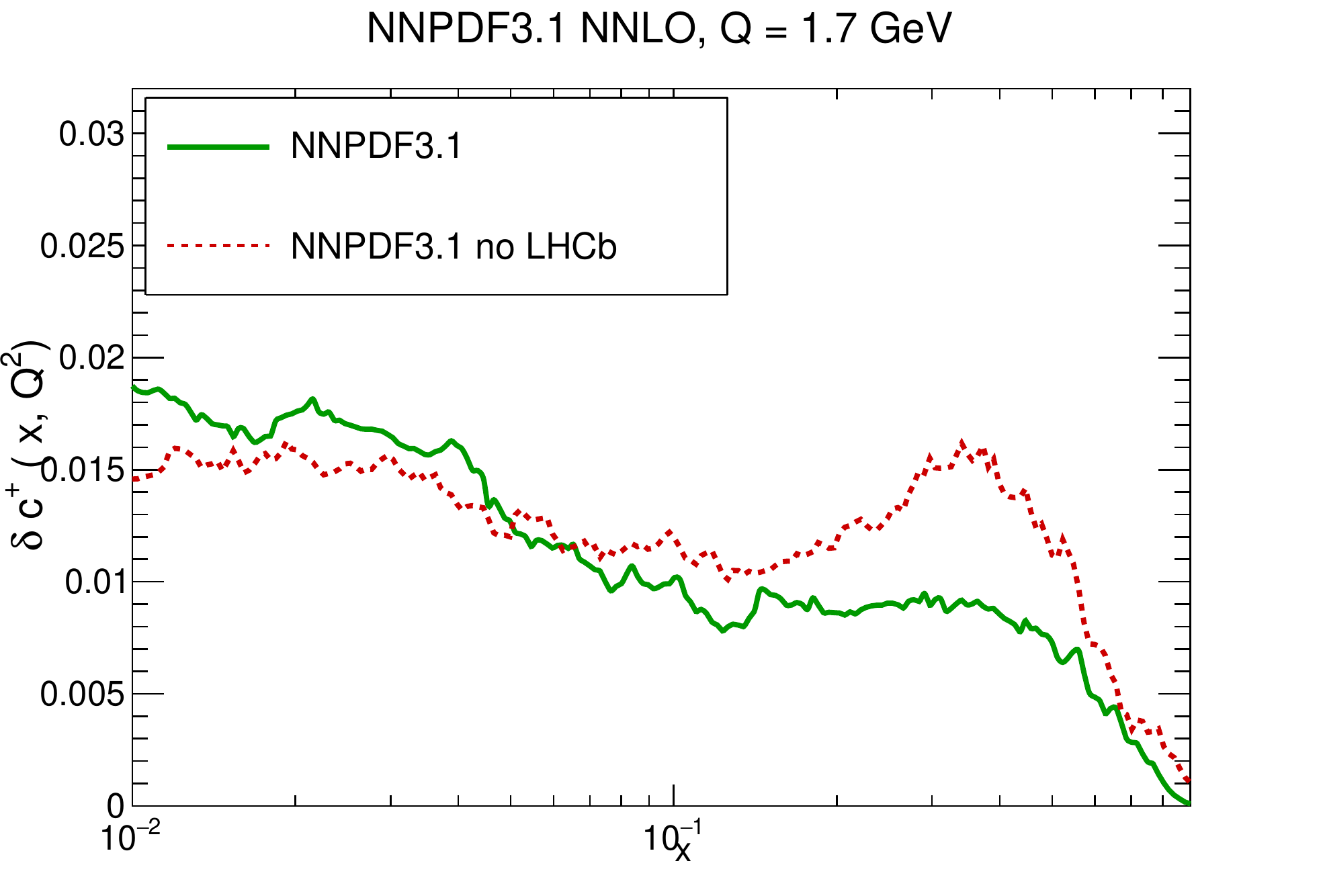}
\caption{\small
  Left plot: the charm PDF $xc^+(x,Q^2)$ at $Q=1.7$ GeV, comparing
  the results of the NNPDF3.1 fit with those of the same fit excluding
  the LHCb data.
  Right plot: the decrease in the absolute PDF uncertainties of the charm
  PDF at the same scale $Q=1.7$ GeV once the LHCb data is included.
  }
\label{fig:charm}
\end{figure}

In Fig.~\ref{fig:charm} we show the
charm PDF $xc^+(x,Q^2)$ at $Q=1.7$ GeV and its
absolute uncertainty $\delta c^+$ for the fits with and without LHCb data.
We observe how the LHCb measurements lead to a suppressed $xc^+$
at large-$x$, as well as to a reduction of the associated
PDF uncertainties.
These results indicate that stringent constraints on models of
the
non-perturbative charm content of the proton can be
provided by the LHCb $W,Z$ data.

%%%%%%%%%%%%%%%%%%%%%%%%%%%%%%%%%%%%%%%%%%%%%%%%%%%%%%%%%%%%%
%%%%%%%%%%%%%%%%%%%%%%%%%%%%%%%%%%%%%%%%%%%%%%%%%%%%%%%%%%%%%%%%%%

The impact of the LHCb data
on the charm PDF can also be gauged by computing $\la xc\ra$, the average
momentum fraction carried by charm quarks in the proton, for
the NNPDF3.1 fits
with and without the LHCb data.
When LHCb data is excluded we find that, for $Q=Q_0=1.65$ GeV,
$\la xc\ra_{\rm noLHCb} =0.012 \pm 0.006$,
while in the baseline NNPDF3.1 fit we have
instead $\la xc\ra_{\rm 3.1} =0.004 \pm 0.004$.
This is a consequence that, as shown in Fig.~\ref{fig:charm},
both the central value is
reduced and the PDF error decreases.
It is also interesting to note that, after the addition of
the EMC charm data (which is not part
of the NNPDF3.1 dataset),
one finds $\la xc\ra_{3.1+EMC}=0.005 \pm 0.001$, a central value
consistent with the one preferred by LHCb.
These results confirm the importance of the forward $W,Z$ LHCb data for our
understanding of the charm content of the nucleon.

\vspace{0.3cm}
\noindent
{\bf Acknowledgments.}
We are grateful to S.~Farry, P.~Ilten and  R.~McNulty
for discussions about the LHCb $W$ and $Z$ data.
This work has been supported
by the ERC Starting Grant ``PDF4BSM''.

\providecommand{\href}[2]{#2}\begingroup\raggedright\endgroup

\end{document}